\documentclass[12pt]{article}
\usepackage{amssymb}
\usepackage{graphicx}
\begin{document}
\newcommand{\beq}{\begin{equation}}
\newcommand{\eeq}{\end{equation}}
\newcommand{\beqa}{\begin{eqnarray}}
\newcommand{\eeqa}{\end{eqnarray}}
\newcommand{\beqar}{\begin{eqnarray*}}
\newcommand{\eeqar}{\end{eqnarray*}}
\newcommand{\al}{\alpha}
\newcommand{\be}{\beta}
\newcommand{\del}{\delta}
\newcommand{\D}{\Delta}
\newcommand{\eps}{\epsilon}
\newcommand{\ga}{\gamma}
\newcommand{\Ga}{\Gamma}
\newcommand{\ka}{\kappa}
\newcommand{\nn}{\nonumber}
\newcommand{\inn}{\!\cdot\!}
\newcommand{\h}{\eta}
\newcommand{\ii}{\iota}
\newcommand{\kk}{\varphi}
\newcommand\F{{}_3F_2}
\newcommand{\la}{\lambda}
\newcommand{\La}{\Lambda}
\newcommand{\na}{\prt}
\newcommand{\Om}{\Omega}
\newcommand{\om}{\omega}
\newcommand{\p}{\Phi}
\newcommand{\sig}{\sigma}
\renewcommand{\t}{\theta}
\newcommand{\z}{\zeta}
\newcommand{\ssc}{\scriptscriptstyle}
\newcommand{\eg}{{\it e.g.,}\ }
\newcommand{\ie}{{\it i.e.,}\ }
\newcommand{\labell}[1]{\label{#1}} %{\label{#1}} %
\newcommand{\reef}[1]{(\ref{#1})}
\newcommand\prt{\partial}
\newcommand\veps{\varepsilon}
\newcommand{\pol}{\varepsilon}
\newcommand\vp{\varphi}
\newcommand\ls{\ell_s}
\newcommand\cF{{\cal F}}
\newcommand\cA{{\cal A}}
\newcommand\cS{{\cal S}}
\newcommand\cT{{\cal T}}
\newcommand\cV{{\cal V}}
\newcommand\cL{{\cal L}}
\newcommand\cM{{\cal M}}
\newcommand\cN{{\cal N}}
\newcommand\cG{{\cal G}}
\newcommand\cK{{\cal K}}
\newcommand\cH{{\cal H}}
\newcommand\cI{{\cal I}}
\newcommand\cJ{{\cal J}}
\newcommand\cl{{\iota}}
\newcommand\cP{{\cal P}}
\newcommand\cQ{{\cal Q}}
\newcommand\cg{{\it g}}
\newcommand\cR{{\cal R}}
\newcommand\cB{{\cal B}}
\newcommand\cO{{\cal O}}
\newcommand\tcO{{\tilde {{\cal O}}}}
\newcommand\bz{\bar{z}}
\newcommand\bb{\bar{b}}
\newcommand\ba{\bar{a}}
\newcommand\bg{\bar{g}}
\newcommand\bc{\bar{c}}
\newcommand\bw{\bar{w}}
\newcommand\bX{\bar{X}}
\newcommand\bK{\bar{K}}
\newcommand\bA{\bar{A}}
\newcommand\bZ{\bar{Z}}
\newcommand\bxi{\bar{\xi}}
\newcommand\bphi{\bar{\phi}}
\newcommand\bpsi{\bar{\psi}}
\newcommand\bprt{\bar{\prt}}
\newcommand\bet{\bar{\eta}}
\newcommand\btau{\bar{\tau}}
\newcommand\hF{\hat{F}}
\newcommand\hA{\hat{A}}
\newcommand\hT{\hat{T}}
\newcommand\htau{\hat{\tau}}
\newcommand\hD{\hat{D}}
\newcommand\hf{\hat{f}}
\newcommand\hK{\hat{K}}
\newcommand\hg{\hat{g}}
\newcommand\hp{\hat{\Phi}}
\newcommand\hi{\hat{i}}
\newcommand\ha{\hat{a}}
\newcommand\hb{\hat{b}}
\newcommand\hQ{\hat{Q}}
\newcommand\hP{\hat{\Phi}}
\newcommand\hS{\hat{S}}
\newcommand\hX{\hat{X}}
\newcommand\tL{\tilde{\cal L}}
\newcommand\hL{\hat{\cal L}}
\newcommand\MZ{\mathbb{Z}}
\newcommand\tG{{\tilde G}}
\newcommand\tg{{\tilde g}}
\newcommand\tphi{{\widetilde \Phi}}
\newcommand\tPhi{{\widetilde \Phi}}
\newcommand\te{{\tilde e}}
\newcommand\tk{{\tilde k}}
\newcommand\tf{{\tilde f}}
\newcommand\ta{{\tilde a}}
\newcommand\tb{{\tilde b}}
\newcommand\tc{{\tilde c}}
\newcommand\td{{\tilde d}}
\newcommand\tm{{\tilde m}}
\newcommand\tmu{{\tilde \mu}}
\newcommand\tnu{{\tilde \nu}}
\newcommand\talpha{{\tilde \alpha}}
\newcommand\tbeta{{\tilde \beta}}
\newcommand\trho{{\tilde \rho}}
 \newcommand\tR{{\tilde R}}
\newcommand\teta{{\tilde \eta}}
\newcommand\tF{{\widetilde F}}
\newcommand\tK{{\widetilde K}}
\newcommand\tE{{\tilde E}}
\newcommand\tpsi{{\tilde \psi}}
\newcommand\tX{{\widetilde X}}
\newcommand\tD{{\widetilde D}}
\newcommand\tO{{\widetilde O}}
\newcommand\tS{{\tilde S}}
\newcommand\tB{{\tilde B}}
\newcommand\tA{{\widetilde A}}
\newcommand\tT{{\widetilde T}}
\newcommand\tC{{\widetilde C}}
\newcommand\tV{{\widetilde V}}
\newcommand\thF{{\widetilde {\hat {F}}}}
\newcommand\Tr{{\rm Tr}}
\newcommand\tr{{\rm tr}}
\newcommand\STr{{\rm STr}}
\newcommand\hR{\hat{R}}
\newcommand\M[2]{M^{#1}{}_{#2}}

\newcommand\bS{\textbf{ S}}
\newcommand\bI{\textbf{ I}}
\newcommand\bJ{\textbf{ J}}

%\begin{document}
\begin{titlepage}
\begin{center}

\vskip 2 cm
{\LARGE \bf More on  $\beta$-symmetry
 }\\
\vskip 1.25 cm
  Mohammad R. Garousi\footnote{garousi@um.ac.ir}

\vskip 1 cm
{{\it Department of Physics, Faculty of Science, Ferdowsi University of Mashhad\\}{\it P.O. Box 1436, Mashhad, Iran}\\}
\vskip .1 cm
 \end{center}

\begin{abstract}

Recent work has proposed a method for imposing T-duality on the metric, $B$-field, and dilaton of the classical effective action of string theory without using Kaluza-Klein reduction. Specifically, the $D$-dimensional effective action should be invariant under global $O(D,D)$ transformations, provided that the partial derivatives along the $\beta$-parameters of the non-geometrical elements of the $O(D,D)$ group become zero. In this paper, we speculate that the global $\beta$-symmetry can be utilized to identify both bulk and boundary couplings.

We demonstrate that the Gibbons-Hawking term at the leading order of $\alpha'$ is reproduced by this symmetry. Additionally, for closed spacetime manifolds at order $\alpha'$, we show that the 20 independent geometrical couplings at this order are fixed by this symmetry up to field redefinitions. Specifically, we show that the invariance of the most general covariant and gauge invariant bulk action at order $\alpha'$ under the most general covariant deformed $\beta$-transformation at order $\alpha'$ fixes  the action  up to 13 parameters. These parameters reflect the field redefinitions freedom for the closed spacetime manifolds. For particular values of these parameters, we recover the effective action in the Metsaev-Tseytlin and in the Meissner schemes.
\end{abstract}

%Keywords: T-duality, D-brane effective action
\end{titlepage}

\section{Introduction}

The classical effective action of string theory and its non-perturbative objects can be determined by imposing gauge symmetries and T-duality. This duality was first observed in the spectrum of string theory when compactified on a circle \cite{Giveon:1994fu,Alvarez:1994dn}. It has been shown in \cite{Sen:1991zi,Hohm:2014sxa} that the Kaluza-Klein reduction of the classical spacetime effective action of string theory on a torus $T^{(d)}$ is invariant under rigid $O(d,d)$-transformations at all orders of $\alpha'$. 
However, the non-perturbative objects D$_p$-brane/O$_p$-plane are not invariant under the rigid $O(d,d)$-transformations, but they are transformed covariantly under a $\MZ_2$-subgroup of the $O(d,d)$-group \cite{Bergshoeff:1996cy}. By assuming that the effective actions at the critical dimension are background independent, one can consider a particular background that includes one circle. Then, one can use the non-geometrical $\MZ_2$-subgroup of the rigid $O(1,1)$-group to construct the higher-derivative corrections to the bulk and boundary spacetime effective actions of string theory \cite{Garousi:2019wgz,Garousi:2019mca,Garousi:2019xlf,Razaghian:2018svg,Garousi:2020gio,Garousi:2020lof,Garousi:2021yyd,Garousi:2021cfc,Garousi:2022ovo,Garousi:2022ghs}. 
Furthermore, it has been shown in \cite{Garousi:2013gea,Robbins:2014ara,Garousi:2014oya,Akou:2020mxx,Mashhadi:2020mzf,Hosseini:2022vrr} that the $\MZ_2$-constraint can be used to construct the higher-derivative corrections to the bulk and boundary D$_p$-brane/O$_p$-plane actions. The $\MZ_2$-transformations in the base space are the Buscher rules \cite{Buscher:1987sk,Buscher:1987qj}, plus their $\alpha'$-corrections \cite{Kaloper:1997ux} that depend on the scheme of the effective action \cite{Garousi:2019wgz}. The constant parameter of the $O(1,1)$-group corresponds to the trivial geometrical re-scaling transformation that receives no $\alpha'$-corrections.

The rigid $O(d,d)$-group contains geometrical subgroups consisting of rigid diffeomorphisms with $d^2$ parameters and shifts on the $B$-field with $d(d-1)/2$ parameters, as well as the following non-geometrical $\beta$-transformations:
\beqa
\delta E_{\mu\nu}&=&-E_{\mu\rho}\beta^{\rho\sigma}E_{\sigma\nu}\,,\nn\\
\delta\Phi&=&\frac{1}{2}E_{\mu\nu}\beta^{\mu\nu}\,.\labell{b-trans}
\eeqa
Here, $\Phi$ is the dilaton, $E_{\mu\nu}=G_{\mu\nu}+B_{\mu\nu}$, and $\beta^{\mu\nu}$ is a constant antisymmetric bi-vector with $d(d-1)/2$ parameters \cite{Maharana:1992my}. A diffeomorphism invariant and $B$-field gauge invariant effective action at the critical dimension $D$ is naturally invariant under the geometrical subgroups of the $O(D,D)$-group with non-constant parameters. The invariance of the $D$-dimensional covariant action under the $\beta$-transformations then demands that the effective action is invariant under the rigid $O(D,D)$-transformations.
Recently, in \cite{Baron:2022but}, it has been proposed that the universal sector of the effective actions of string theories at any order of $\alpha'$ should be invariant under the $\beta$-transformations with the following constraint on the partial derivatives:
\beqa
\beta^{\mu\nu}\prt_{\nu}(\cdots)&=&0 \,.\labell{b-cons}
\eeqa
The $\beta$-transformations \reef{b-trans} should be deformed at higher orders of $\alpha'$, as in the Buscher rules.

Using the frame formalism, in which the frames $e^a{}_\mu$ are defined as $e^a{}_\mu e^b{}_\nu\eta_{ab}=G_{\mu\nu}$ and are the independent fields, it has been shown in \cite{Baron:2022but} that the following effective action at the leading order of $\alpha'$ is invariant under the $\beta$-transformations:
\beqa
  \bS^0= -\frac{2}{\kappa^2}\int d^{D} x\sqrt{-G} e^{-2\Phi} \left(  R - 4\nabla_{\mu}\Phi \nabla^{\mu}\Phi+4\nabla_\mu\nabla^\mu\Phi-\frac{1}{12}H^2\right)\,.\labell{S0bf}
\eeqa
Here, $H$ is the field strength of the $B$-field. The existence of this symmetry at the supergravity level was already noted in \cite{Andriot:2014uda}. It has been observed in \cite{Baron:2022but} that the $\beta$-transformations \reef{b-trans} in combination with the local Lorentz transformations satisfy a closed algebra. It has also been shown that the effective action at order $\alpha'$ in a particular scheme, namely the two-parameter generalized Bergshoeff-de Roo scheme, in which the local Lorentz transformations and the $B$-field gauge transformations receive a particular $\alpha'$-correction \cite{Marques:2015vua}, is invariant under a deformed $\beta$-transformations. The corresponding deformations at order $\alpha'$ for the $\beta$-transformations \reef{b-trans} have been found. It has also been shown that the deformed local transformations and the deformed $\beta$-transformations satisfy a closed algebra \cite{Baron:2022but}.

Assuming that the effective actions of string theory at the critical dimension are background-independent \cite{Garousi:2022ovo}, we can expect that the global symmetry observed in the effective action of closed spacetime manifolds also applies to open spacetime manifolds. In the latter case, we have both bulk and boundary actions, denoted as $S$ and $\partial S$ respectively.
In certain schemes, the sum of these two actions may exhibit invariance under global transformations, represented as:
\beqa
\delta(S+\partial S) &=& 0 \,.\labell{sym}
\eeqa
In other schemes, each action may possess separate invariance:
\beqa
\delta S = 0 \quad ; \quad \delta(\partial S) = 0\,. \labell{sym1}
\eeqa
In the former cases, the bulk action remains invariant under the global symmetry, albeit with residual total derivative terms. Similarly, the boundary action is also invariant, but may contain anomalous terms. However, these anomalous terms can be canceled out by the residual total derivative terms in the bulk through the utilization of Stokes' theorem.
In the latter cases, both the bulk and boundary actions lack residual total derivative terms and anomalous terms respectively. In this paper, we are going to study  bulk and boundary covariant couplings by imposing invariance under the $\beta$-transformations.

To determine the bulk and boundary actions with $\beta$-symmetry at any order of $\alpha'$ in an arbitrary scheme, one must consider the most general bulk and boundary covariant and gauge invariant couplings, excluding total derivative terms, with arbitrary coefficients. The next step involves fixing the parameters by ensuring the actions remain invariant under the deformed $\beta$-transformations and verifying that the combined deformed $\beta$-transformations and standard local transformations satisfy a closed algebra.
However, in this paper, our focus is solely on imposing the invariance of effective actions and determining the extent to which this constraint can fix the parameters within the bulk effective action.
Such a calculation in bosonic string theory at order $\alpha'$ has been performed in \cite{Garousi:2019wgz} by imposing the invariance of the action under the deformed Buscher rules.

The outline of the paper is as follows: 
In section 2, we demonstrate that in the presence of a boundary, the combination of the leading-order bulk effective action  and the Gibbons-Hawking boundary term satisfies the $\beta$-symmetry.
In section 3, we investigate the spacetime effective action at order $\alpha'$. We assume no deformation for the local transformations and use the most general covariant deformations at order $\alpha'$ for the $\beta$-transformations given in Eq.~(\ref{b-trans}). We show that the invariance of the most general bulk couplings under the $\beta$-transformations fixes the couplings  up to 13 parameters. The couplings are exactly the same as those found by the deformed Buscher rules.
In section 4, we provide a brief discussion of our results. We have used the "xAct" package \cite{Nutma:2013zea} for performing our calculations in this paper.

\section{The $\beta$-symmetry at leading order}

In \cite{Baron:2022but}, it was shown that the leading-order bulk action given in Eq.~(\ref{S0bf}) is completely invariant under the $\beta$-transformation, meaning that there are no residual total derivative terms. However, this action is not the standard spacetime effective action of string theory. For closed spacetime manifolds, where there is no boundary, one can use an integration by parts to convert the effective action to the standard form, which has no Laplacian of the dilaton, as 
\beqa
  \bS^0= -\frac{2}{\kappa^2}\int d^{D} x\sqrt{-G} e^{-2\Phi} \left(  R + 4\nabla_{\mu}\Phi \nabla^{\mu}\Phi-\frac{1}{12}H^2\right)\,.\labell{S0bf1}
\eeqa
However, for spacetime manifolds with boundaries, one expects both boundary and bulk couplings. At the leading order of $\alpha'$, there is the Gibbons-Hawking boundary term for the spacetime action \cite{Gibbons:1976ue}. However, as we will see shortly, this term is not invariant under the $\beta$-transformations. Therefore, the $\beta$-symmetry does not allow us to add the Gibbons-Hawking term to the bulk action given in Eq.~(\ref{S0bf}).
In \cite{Garousi:2019xlf}, it was shown that if one adds the Gibbons-Hawking term to the standard form of the bulk action given in Eq.~(\ref{S0bf1}), then the resulting action is invariant under the Buscher rules. Therefore, we expect this combination to be invariant under the $\beta$-transformations as well.

We consider a spacetime $M^D$ with a boundary $\partial M^D$. Using the transformations given in Eq.~(\ref{b-trans}) and the constraint given in Eq.~(\ref{b-cons}), one can derive the following transformations for the different terms in Eq.~(\ref{S0bf1}):
\beqa
\delta(\sqrt{-G} e^{-2\Phi})&=& \sqrt{-G} e^{-2\Phi}\bigg[-2\delta\Phi+\frac{1}{2}G^{\mu\nu}\delta G_{\nu\mu}\bigg]\,=\,0\,,\nn\\
\delta(R)&=&-2G^{\ga\delta}\nabla_\ga\nabla_{\delta}(B_{\mu\nu}\beta^{\mu\nu})+G^{\alpha\ga}G^{\beta\delta}\nabla_{[\ga} (B_{\delta] \mu}\beta_1^{\mu})\nabla_{[\alpha}(G_{\beta]\nu}\beta_2^{\nu})-(\beta_1\leftrightarrow\beta_2)\,,\nn\\
\delta(\nabla_{\mu}\Phi \nabla^{\mu}\Phi)&=&G^{\alpha\beta}\nabla_{\alpha}(B_{\mu\nu}\beta^{\mu\nu})\nabla_{\beta}\Phi\,,\nn\\
\delta(H^2)&=&12G^{\alpha\ga}G^{\beta\delta}\nabla_{[\ga} (B_{\delta] \mu}\beta_1^{\mu})\nabla_{[\alpha}(G_{\beta]\nu}\beta_2^{\nu})-(\beta_1\leftrightarrow\beta_2)\,,\labell{trans}
\eeqa
where we have also defined the vectors $\beta_1^\mu, \beta_2^\mu$ as\footnote{It is worth noting that the terms on the right-hand side of the second and fourth lines in equation \reef{trans} can be expressed in terms of $\beta^{\mu\nu}$. As a result, the combination $(\beta_1^\mu\beta_2^\nu+\beta_1^\nu\beta_2^\mu)$, which encompasses $d(d+1)/2$ parameters, does not appear in the transformations stated in equation \reef{trans}. Therefore, only $d(d-1)/2$ parameters of $\beta_1^\mu$ and $\beta_2^\mu$ appear in the transformations, precisely the same parameters that $\beta^{\mu\nu}$ possesses.}
\beqa
\beta^{\mu\nu}=\frac{1}{2}(\beta_1^\mu\beta_2^\nu-\beta_1^\nu\beta_2^\mu)\,.
\eeqa
It is worth noting that the field $(B_{\mu\nu}\beta^{\mu\nu})$ is a scalar in spacetime. Similarly, the fields $G_{\mu\alpha}\beta_1^\alpha$, $G_{\mu\alpha}\beta_2^\alpha$, $B_{\mu\alpha}\beta_1^\alpha$, and $B_{\mu\alpha}\beta_2^\alpha$ are vectors in spacetime. Therefore, the right-hand side terms in Eq.~(\ref{trans}) are covariant and invariant under the gauge transformations corresponding to these vectors.

Upon substituting the transformations given in Eq.~(\ref{trans}) into the action given in Eq.~(\ref{S0bf1}), one can observe that the transformation of $H^2$ is canceled by the last two terms in the transformation of $R$. Similarly, the transformation of $(\nabla\Phi)^2$ combines with the first term in the transformation of $R$ to produce the following covariant derivative:
\beqa
\delta(\!\!\bS^0)=\frac{4}{\kappa^2}\int d^{D} x\sqrt{-G}\, \nabla^\ga\,J_\ga\,,\labell{bound0}
\eeqa
where the  Noether current is
\beqa
J_\ga&=& e^{-2\Phi}H_{\ga\mu\nu}\beta^{\mu\nu}\,.
\eeqa
The presence of the total derivative term given in Eq.~(\ref{bound0}) indicates that for spacetime manifolds with a boundary, there must be some covariant couplings on the boundary that are not invariant under the $\beta$-transformations either.
Using Stokes' theorem, one can write the anomalous term given in Eq.~(\ref{bound0}) as:
\beqa
\delta(\!\!\bS^0)=\frac{4}{\kappa^2}\int d^{D-1}\sigma\sqrt{|g|}\, n^\gamma\bigg[e^{-2\Phi}H_{\gamma\mu\nu}\beta^{\mu\nu}\bigg]\,.\labell{bound}
\eeqa
Here, $n_\mu$ is the outward-pointing normal vector to the boundary, which is spacelike (timelike) if the boundary is spacelike (timelike). Inside the square root, $g$ is the determinant of the induced metric on the boundary, which is given by:
\beqa
g_{\tilde{\mu}\tilde{\nu}}=\frac{\partial x^\mu}{\partial\sigma^{\tilde{\mu}}}\frac{\partial x^\nu}{\partial\sigma^{\tilde{\nu}}}G_{\mu\nu}\,.\labell{indg}
\eeqa
Here, the boundary is specified by the functions $x^\mu=x^\mu(\sigma^{\tilde{\mu}})$, where $\sigma^\tmu$ are the coordinates of the boundary $\partial M^D$.

Let us now consider the following boundary term:
\beqa
\prt\!\!\bS^0=-\frac{2a}{\kappa^2}\int d^{D-1}\sigma\sqrt{|g|}\, e^{-2\Phi}G^{\mu\nu}K_{\mu\nu}\,.\labell{bound1}
\eeqa
Here, $a$ is a constant parameter, and $K_{\mu\nu}$ is the extrinsic curvature given by:
\beqa
K_{\mu\nu}=\nabla_\mu n_\nu\mp n_\mu n^\rho\nabla_\rho n_\nu\,.
\eeqa
The minus (plus) sign is used for timelike (spacelike) boundaries. We specify the boundary as $x^\mu(\sigma^\tmu)=(\sigma^\tmu,z_*)$, where $z_*$ is fixed on the boundary. The normal vector becomes along the $z$-direction, and the induced metric \reef{indg} becomes $g_{\tmu\tnu}=G_{\tmu\tnu}$. The measure in \reef{bound1} becomes invariant under the $\beta$-transformations:
\beqa
\delta(\sqrt{|g|} e^{-2\Phi})&=& \sqrt{|g|} e^{-2\Phi}\bigg[-2\delta\Phi+\frac{1}{2}G^{\tmu\tnu}\delta G_{\tnu\tmu}\bigg]\,=\,0\,.
\eeqa
Here, $G^{\tmu\tnu}$ is the inverse of the induced metric $G_{\tmu\tnu}$, and we have used the $\beta$-transformations of metric and dilaton:
\beqa
\delta G_{\tmu\tnu}&=&-G_{\tmu \talpha}\beta^{\talpha\tbeta}B_{\tbeta\tnu}-G_{\tnu \talpha}\beta^{\talpha\tbeta}B_{\tbeta\tmu}\,,\nn\\
\delta\Phi &=&\frac{1}{2}B_{\tmu\tnu}\beta^{\tmu\tnu}\,.
\eeqa
These transformations are the same as the transformation \reef{b-trans}, in which the following constraint is also imposed:
\beqa
n_\mu\beta^{\mu\nu}=\beta^{z\nu}&=&0\,.\labell{nb-cons}
\eeqa
This latter constraint is similar to the $\mathbb{Z}_2$-constraint for circular reduction, in which it is assumed that the unit vector $n_\mu$ has no component along the killing circle \cite{Garousi:2022ovo}. The transformation of the boundary action \reef{bound1} under the $\beta$-transformations \reef{b-trans} and the constraints \reef{b-cons} and \reef{nb-cons} becomes:
\beqa
\delta(\prt\!\!\bS^0)&=&-\frac{2a}{\kappa^2}\int d^{d-1} \sigma\sqrt{|g|}\, e^{-2\Phi} n^\gamma H_{\gamma\mu\nu}\beta^{\mu\nu}\,.\labell{bound2}
\eeqa
The above transformation cancels the anomalous term in \reef{bound} for $a=2$. Hence, the following bulk and boundary actions:
\beqa
\bS^0+\prt\!\!\bS^0
=-\frac{2}{\kappa^2}\Bigg[ \int d^{D}x \sqrt{-G} e^{-2\Phi} \left(  R + 4\nabla_{\mu}\Phi \nabla^{\mu}\Phi-\frac{1}{12}H^2\right)+ 2\int d^{D-1}\sigma\sqrt{| g|}  e^{-2\Phi}K\Bigg],\labell{baction}
\eeqa
are invariant under the $\beta$-transformations, i.e., they satisfy \reef{sym}. Note that the coefficient of the boundary term is exactly the one that appears in the Gibbons-Hawking boundary term \cite{Gibbons:1976ue}. 

It is worth noting that we could also include the term $n^\alpha\nabla_\alpha\Phi$ in the boundary coupling \reef{bound1}. Such a term would also be consistent with the $\beta$-symmetry; however, it would not be consistent with the least action principle \cite{Garousi:2021cfc} if one considers \reef{S0bf1} as the bulk action. If we consider the action \reef{S0bf} as the bulk action, then we can include the term $n^\alpha\nabla_\alpha\Phi$ in the boundary action. In fact, the following combination is invariant under the $\beta$-transformation:
\beqa
\partial\!\!\bS^{(0)}&=&  -\frac{2a_1}{\kappa^2}\int d^{D-1}\sigma\, e^{-2\Phi}\sqrt{|g|}\,  \left(-\frac{1}{2} K+n^{\mu}\nabla_{\mu}\Phi \right)\,.\labell{bbaction}
\eeqa
The above boundary couplings are consistent with the least action principle when $a_1=-4$. The sum of the bulk action \reef{S0bf} and the above boundary action, after applying Stokes' theorem, becomes \reef{baction}.

%Although the above calculation was straightforward at the leading order of $\alpha'$, extending it to higher orders is non-trivial. This is because finding both bulk and boundary actions by the $\beta$-symmetry may require deforming the $\beta$-transformations as well as the local transformations at higher orders of $\alpha'$.

In the upcoming section, our attention will be directed towards the bulk couplings at order $\alpha'$ for closed spacetime manifolds. We will proceed under the assumption that these bulk couplings remain invariant under the standard local transformations while permitting covariant deformations of the $\beta$-transformations. By adopting this assumption, we will discover that the invariance of the bulk action under the deformed $\beta$-transformations serves to determine the bulk couplings, up to the most general field redefinitions.

\section{The $\beta$-symmetry at order $\alpha'$}

If the local Lorentz and diffeomorphism and $B$-field gauge transformations are not deformed at order $\alpha'$, then the most general bulk couplings involve 41 covariant and gauge-invariant terms. However, some of these terms are related by total derivative terms, while others are related by the Bianchi identities. Removing these redundancies, one can identify 20 independent even-parity couplings \cite{Metsaev:1987zx}. That is 
\beqa
 \bS^1&=& -\frac{2\alpha'}{\kappa^2}\int d^{D} x\sqrt{-G} e^{-2\Phi} \Bigg[   a_{1}^{} H_{\alpha }{}^{\delta \epsilon }
H^{\alpha \beta \gamma } H_{\beta \delta }{}^{\varepsilon } H_{
\gamma \epsilon \varepsilon } + a_{2}^{} H_{\alpha \beta
}{}^{\delta } H^{\alpha \beta \gamma } H_{\gamma }{}^{\epsilon
\varepsilon } H_{\delta \epsilon \varepsilon }\nn\\&& + a_{3}^{}
H_{\alpha \beta \gamma } H^{\alpha \beta \gamma } H_{\delta
\epsilon \varepsilon } H^{\delta \epsilon \varepsilon } +
a_{4}^{} H_{\alpha }{}^{\gamma \delta } H_{\beta \gamma \delta
} R^{\alpha \beta } + a_{5}^{} R_{\alpha \beta } R^{\alpha
\beta } + a_{8}^{} R_{\alpha \beta \gamma \delta } R^{\alpha
\beta \gamma \delta } \nn\\&&+ a_{9}^{} H_{\alpha }{}^{\delta
\epsilon } H^{\alpha \beta \gamma } R_{\beta \gamma \delta
\epsilon } + a_{11}^{} R \nabla_{\alpha }\nabla^{\alpha }\Phi
+ a_{10}^{} H_{\beta \gamma \delta } H^{\beta \gamma \delta }
\nabla_{\alpha }\nabla^{\alpha }\Phi + a_{13}^{} R
\nabla_{\alpha }\Phi \nabla^{\alpha }\Phi \nn\\&&+ a_{12}^{}
H_{\beta \gamma \delta } H^{\beta \gamma \delta }
\nabla_{\alpha }\Phi \nabla^{\alpha }\Phi + a_{14}^{}
\nabla_{\alpha }\Phi \nabla^{\alpha }\Phi \nabla_{\beta
}\nabla^{\beta }\Phi + a_{15}^{} H_{\alpha }{}^{\gamma \delta
} H_{\beta \gamma \delta } \nabla^{\alpha }\Phi \nabla^{\beta
}\Phi\nn\\&& + a_{16}^{} R_{\alpha \beta } \nabla^{\alpha }\Phi
\nabla^{\beta }\Phi + a_{17}^{} \nabla_{\alpha }\Phi \nabla^{
\alpha }\Phi \nabla_{\beta }\Phi \nabla^{\beta }\Phi +
a_{18}^{} H_{\alpha }{}^{\gamma \delta } H_{\beta \gamma
\delta } \nabla^{\beta }\nabla^{\alpha }\Phi \nn\\&&+ a_{19}^{}
\nabla_{\beta }\nabla_{\alpha }\Phi \nabla^{\beta
}\nabla^{\alpha }\Phi + a_{20}^{} \nabla_{\alpha }H^{\alpha
\beta \gamma } \nabla_{\delta }H_{\beta \gamma }{}^{\delta }+ a_{7}^{} R^2 + a_{6}^{} R H_{\alpha \beta \gamma } H^{\alpha
\beta \gamma } \Bigg]\,, \labell{L1bulk}
\eeqa
 where $a_1,\cdots, a_{20}$ are 20 parameters that the gauge symmetries can not fix them.

%Assuming the most general deformation for the Buscher rules at order $\alpha'$, the $\mathbb{Z}_2$-symmetry produces seven relations between the parameters in the bulk couplings and fixes the corrections to the Buscher rules in terms of the remaining parameters \cite{Garousi:2019wgz}. The unfixed parameters are arbitrary, but for some relations between them, the deformation does not ruin the data on the boundary at order $\alpha'$. In this subset of parameters, one can also determine the boundary action \cite{Garousi:2021yyd,Garousi:2021cfc}.
%If the most general field redefinitions are used in the bulk action, the couplings in \reef{L1bulk} can be reduced to eight independent couplings \cite{Metsaev:1987zx}. In this case, the $\mathbb{Z}_2$-constraint fixes all of them, as well as the corrections to the Buscher rules up to an overall factor. However, the corrections to the Buscher rules ruin the data on the boundary \cite{Garousi:2019wgz}, meaning that the boundary terms cannot be represented in the minimal scheme. Therefore, in the presence of a boundary, one cannot use the most general field redefinitions to reduce the bulk couplings in \reef{L1bulk} and still find both the bulk and boundary actions via the $\mathbb{Z}_2$-symmetry.
In this section, we assume the most general covariant deformation for the $\beta$-transformations at order $\alpha'$. By requiring the bulk action to be invariant under the deformed $\beta$-transformations, we can then determine the relations between the parameters in \reef{L1bulk}.
 Specifically, we consider:
\beqa
\delta\Phi&=&\delta\Phi^{(0)}+\alpha' \delta\Phi^{(1)}+\cdots\,,\nn\\
\delta G_{\alpha\beta}&=&\delta G_{\alpha\beta}^{(0)}+\alpha' \delta G_{\alpha\beta}^{(1)}+\cdots\,,\nn\\
\delta B_{\alpha\beta}&=&\delta B_{\alpha\beta}^{(0)}+\alpha' \delta B_{\alpha\beta}^{(1)}+\cdots\,,\labell{deform}
\eeqa
where $\delta\Phi^{(0)}$,  $\delta G_{\alpha\beta}^{(0)}$,  $\delta B_{\alpha\beta}^{(0)}$ are given in \reef{b-trans} and 
\beqa
\delta\Phi^{(1)}&=&f_{3} H_{\alpha  \beta  \gamma  } \beta^{\beta  
\gamma  } \nabla^{\alpha  }\Phi  +
 f_{1} \beta^{\alpha  \beta  } \nabla_{\gamma  
}H_{\alpha  \beta  }{}^{\gamma  } +
 f_{2} H_{\alpha  \beta  \gamma  } \nabla^{\gamma  
}\beta^{\alpha  \beta  }\,,\nn\\
\delta G_{\alpha\beta}^{(1)}&=&g_{1}   (\beta^{\gamma  \delta  } 
\nabla_{\alpha  }H_{\beta  \gamma  \delta  } + \beta^{\gamma 
 \delta  } \nabla_{\beta  }H_{\alpha  \gamma  \delta  
}) +   
 g_{4}   (H_{\beta  \gamma  \delta  } 
\nabla_{\alpha  }\beta^{\gamma  \delta  } + H_{\alpha  
\gamma  \delta  } \nabla_{\beta  }\beta^{\gamma  \delta  }
) \nn\\&&+   
 g_{7}   (H_{\beta  \gamma  \delta  } \beta ^{\gamma  \delta  } \nabla_{\alpha  }\Phi  + H_{\alpha  
\gamma  \delta  } \beta^{\gamma  \delta  } \nabla_{\beta  
}\Phi ) +   
 g_{8}   H_{\gamma  \delta  \epsilon  } G_{\alpha  \beta  
} \beta^{\delta  \epsilon  } \nabla^{\gamma  }\Phi \nn\\&& +   
 g_{9}   (H_{\beta  \gamma  \delta  } \beta _{\alpha  }{}^{\delta  } \nabla^{\gamma  }\Phi  + H_{\alpha  
\gamma  \delta  } \beta_{\beta  }{}^{\delta  } 
\nabla^{\gamma  }\Phi ) +   
 g_{2}   (\beta_{\beta  }{}^{\gamma  } 
\nabla_{\delta  }H_{\alpha  \gamma  }{}^{\delta  } + \beta _{\alpha  }{}^{\gamma  } \nabla_{\delta  }H_{\beta  \gamma  
}{}^{\delta  }) \nn\\&&+   
 g_{5}   (H_{\beta  \gamma  \delta  } 
\nabla^{\delta  }\beta_{\alpha  }{}^{\gamma  } + H_{\alpha  
\gamma  \delta  } \nabla^{\delta  }\beta_{\beta  }{}^{\gamma 
 }) +   
 g_{3}   G_{\alpha  \beta  } \beta^{\gamma  \delta  } 
\nabla_{\epsilon  }H_{\gamma  \delta  }{}^{\epsilon  } +   
 g_{6}   H_{\gamma  \delta  \epsilon  } G_{\alpha  \beta  
} \nabla^{\epsilon  }\beta^{\gamma  \delta  }\,,\nn\\
\delta B_{\alpha\beta}^{(1)}&=&e_{5}   H_{\gamma  \delta  \epsilon  } H^{\gamma  \delta  
\epsilon  } \beta_{\alpha  \beta  } +   
 e_{1}   R \beta_{\alpha  \beta  } +   
 e_{2}   (R_{\beta  }{}^{\gamma  } \beta _{\alpha  \gamma  }- R_{\alpha  }{}^{\gamma  } 
\beta_{\beta  \gamma  }) +   
 e_{6}   (H_{\beta  }{}^{\delta  \epsilon  } 
H_{\gamma  \delta  \epsilon  } \beta_{\alpha  }{}^{\gamma  } 
 - H_{\alpha  }{}^{\delta  \epsilon  } H_{\gamma  \delta  
\epsilon  } \beta_{\beta  }{}^{\gamma  }) \nn\\&&+   
 e_{7}   H_{\alpha  \gamma  }{}^{\epsilon  } H_{\beta  
\delta  \epsilon  } \beta^{\gamma  \delta  } +   
 e_{8}   H_{\alpha  \beta  }{}^{\epsilon  } H_{\gamma  
\delta  \epsilon  } \beta^{\gamma  \delta  } +   
 e_{3}   R_{\alpha  \beta  \gamma  \delta  } 
\beta^{\gamma  \delta  } +   
 e_{4}   R_{\alpha  \gamma  \beta  \delta  } 
\beta^{\gamma  \delta  } \nn\\&&+   
 e_{9}   (\nabla_{\alpha  }\nabla_{\gamma  }\beta _{\beta  }{}^{\gamma  }- \nabla_{\beta  }\nabla_{\gamma  
}\beta_{\alpha  }{}^{\gamma  }) +   
 e_{12}   (\nabla_{\beta  }\Phi  \nabla_{\gamma  
}\beta_{\alpha  }{}^{\gamma  }- \nabla_{\alpha  }\Phi  
\nabla_{\gamma  }\beta_{\beta  }{}^{\gamma  }) \nn\\&&+   
 e_{10}   (\nabla_{\gamma  }\nabla_{\alpha  
}\beta_{\beta  }{}^{\gamma  }- \nabla_{\gamma  
}\nabla_{\beta  }\beta_{\alpha  }{}^{\gamma  }) +   
 e_{17}  ( \beta_{\beta  }{}^{\gamma  } \nabla_{
\gamma  }\nabla_{\alpha  }\Phi - \beta_{\alpha  
}{}^{\gamma  } \nabla_{\gamma  }\nabla_{\beta  }\Phi ) 
+  
 e _{11}   \nabla_{\gamma  }\nabla^{\gamma  }\beta_{\alpha 
 \beta  } \nn\\&&+   
 e_{18}   \beta_{\alpha  \beta  } \nabla_{\gamma  
}\nabla^{\gamma  }\Phi  +   
 e_{13}   \nabla_{\gamma  }\beta_{\alpha  \beta  } 
\nabla^{\gamma  }\Phi  +   
 e_{15}   \beta_{\alpha  \beta  } \nabla_{\gamma  
}\Phi  \nabla^{\gamma  }\Phi  +   
 e_{14}   (\nabla_{\alpha  }\beta_{\beta  
\gamma  } \nabla^{\gamma  }\Phi - \nabla_{\beta  }\beta_{
\alpha  \gamma  } \nabla^{\gamma  }\Phi ) \nn\\&&+   
 e_{16} (  \beta_{\beta  \gamma  } 
\nabla_{\alpha  }\Phi  \nabla^{\gamma  }\Phi - \beta _{\alpha  \gamma  } \nabla_{\beta  }\Phi  \nabla^{\gamma  
}\Phi )\,.\labell{delta1}
\eeqa
The parameters $f_1$, $f_2$, $f_3$, $g_1$ through $g_9$, and $e_1$ through $e_{18}$ are all subject to the $\beta$-constraint on the effective action \reef{L1bulk}.  The $\beta$-transformations $\delta\Phi^{(0)}$, $\delta G_{\alpha\beta}^{(0)}$, and $\delta B_{\alpha\beta}^{(0)}$ should be imposed on the action at order $\alpha'$ to produce $\delta(\!\!\bS^1)$. Similarly, $\delta\Phi^{(1)}$, $\delta G_{\alpha\beta}^{(1)}$, and $\delta B_{\alpha\beta}^{(1)}$ should be imposed on the leading-order action \reef{S0bf1} to produce $\alpha'\Delta(\!\!\bS^0)$. The latter produces the following terms at order $\alpha'$:
\beqa
\alpha'\Delta(\!\!\bS^0)&=&-\frac{2\alpha'}{\kappa^2}\int d^{D} x\sqrt{-G} e^{-2\Phi} \Bigg[ - R^{\alpha \beta } \delta G^{(1)}{}_{\alpha \beta } -  \frac{1}{24} H_{\beta \gamma \delta } H^{\beta \gamma \delta } \delta 
G^{(1)\alpha }{}_{\alpha } + \frac{1}{2} R \delta G^{(1)\alpha 
}{}_{\alpha } \nn\\&&+ \frac{1}{4} H_{\alpha }{}^{\gamma \delta } H_{\beta 
\gamma \delta } \delta G^{(1)\alpha \beta } + \frac{1}{6} H_{\alpha 
\beta \gamma } H^{\alpha \beta \gamma } \delta \Phi^{(1)} - 2 R 
\delta \Phi^{(1)} + 8 \nabla_{\alpha }\Phi \nabla^{\alpha }\delta 
\Phi^{(1)} \nn\\&&+ 2 \delta G^{(1)\beta }{}_{\beta } \nabla_{\alpha }\Phi \nabla^{\alpha }\Phi - 8 \delta \Phi^{(1)} \nabla_{\alpha }\Phi \nabla^{\alpha }\Phi + \nabla_{\beta }\nabla_{\alpha }\delta G^{(1)\alpha 
\beta } -  \nabla_{\beta }\nabla^{\beta }\delta G^{(1)\alpha 
}{}_{\alpha } \nn\\&&- 4 \delta G^{(1)}{}_{\alpha \beta } \nabla^{\alpha }\Phi 
\nabla^{\beta }\Phi -  \frac{1}{2} H_{\alpha \beta \gamma } \nabla^{
\gamma }\delta B^{(1)\alpha \beta }\Bigg]\,.\labell{Ds0}
\eeqa
Upon substituting the deformations \reef{delta1} into the equation mentioned above, it becomes evident that certain resulting terms are interconnected through total derivatives, while others may be linked by the Bianchi identities. Furthermore, the deformed $\beta$-transformation \reef{deform}, when combined with the local transformations, must satisfy a closed algebra. As a consequence, the parameters in \reef{delta1} are not all independent. It is possible to determine the relationships between these parameters by imposing the condition of a closed algebra and eliminating the parameters that are related by total derivatives and the Bianchi identities. In such a scenario, the constraint of the effective action's invariance under the resulting deformed $\beta$-transformations could fix both the effective action and the deformed $\beta$-transformations, up to field redefinitions. However, in this paper, our focus lies in fixing the parameters within the effective action \reef{Ds0}. Therefore, we do not impose the aforementioned constraint on the parameters in \reef{delta1}. In general, without imposing the closed algebra constraint, the parameters of the effective action cannot be fixed up to field redefinitions. However, as we will see for the couplings at order $\alpha'$, they are indeed fixed up to field redefinitions.

%One may impose the constraint that the Instead, the independent parameters are fixed by imposing the bulk and boundary constraints that the total derivatives and the deformations must satisfy. After imposing these constraints, any remaining unfixed parameters can be removed from the calculations.
%For open spacetime manifolds, there are more constraints than for closed spacetime manifolds. For example, there is the constraint that the deformations should not spoil the data on the boundary, which does not exist for closed spacetime manifolds.

By using the transformations \reef{b-trans} and the constraint \reef{b-cons}, one can determine the transformation of each coupling in the action \reef{L1bulk}. For example, we have:
\beqa
H_{\alpha }{}^{\delta \epsilon }
H^{\alpha \beta \gamma } H_{\beta \delta }{}^{\varepsilon } H_{
\gamma \epsilon \varepsilon } &\rightarrow& -6 H_{\alpha \gamma }{}^{\epsilon } H_{\beta \delta \epsilon } 
V_1^{\alpha \beta } W_2^{\gamma \delta }-(1\leftrightarrow 2)\,,\nn\\
H_{\alpha \beta
}{}^{\delta } H^{\alpha \beta \gamma } H_{\gamma }{}^{\epsilon
\varepsilon } H_{\delta \epsilon \varepsilon }  &\rightarrow& 4 H_{\beta }{}^{\delta \epsilon } H_{\gamma \delta \epsilon } 
{V_1}^{\alpha \beta } {W_2}_{\alpha }{}^{\gamma } + 2 H_{\alpha \beta 
}{}^{\epsilon } H_{\gamma \delta \epsilon } {V_1}^{\alpha \beta } 
{W_2}^{\gamma \delta }-(1\leftrightarrow 2)\,,\nn\\
H_{\alpha \beta \gamma } H^{\alpha \beta \gamma } H_{\delta
\epsilon \varepsilon } H^{\delta \epsilon \varepsilon }  &\rightarrow& -6 H_{\gamma \delta \epsilon } H^{\gamma \delta \epsilon } 
{V_1}^{\alpha \beta } {W_2}_{\alpha \beta }-(1\leftrightarrow 2)\,,\nn\\
R_{\alpha \beta \gamma \delta } R^{\alpha
\beta \gamma \delta }   &\rightarrow&  \nabla^\alpha V_1^{ \beta \gamma } \nabla_{\alpha}{W_2}_{ \beta \gamma } -  
\frac{3}{2} R_{\alpha \beta \gamma \delta } 
V_1^{\alpha \beta } W_2^{\gamma \delta }-(1\leftrightarrow 2)\,,\nn\\
H_{\alpha }{}^{\delta
\epsilon } H^{\alpha \beta \gamma } R_{\beta \gamma \delta
\epsilon } &\rightarrow&-\nabla^\alpha{V_2}^{ \beta \gamma } H_{\beta \gamma \delta } 
{V_1}_{\alpha }{}^{\delta } + \nabla^\alpha{W_2}^{ \beta \gamma } H_{\beta 
\gamma \delta } {W_1}_{\alpha }{}^{\delta } -  \frac{1}{2} 
H_{\alpha \gamma }{}^{\epsilon } H_{\beta \delta \epsilon } 
{V_1}^{\alpha \beta } {W_2}^{\gamma \delta } \nn\\&&-  \frac{1}{2} 
H_{\alpha \beta }{}^{\epsilon } H_{\gamma \delta \epsilon } 
{V_1}^{\alpha \beta } {W_2}^{\gamma \delta } -  R_{\alpha 
\beta \gamma \delta } {V_1}^{\alpha \beta } {W_2}^{\gamma \delta }-(1\leftrightarrow 2)\,,\nn\\
H_{\beta \gamma \delta } H^{\beta \gamma \delta }
\nabla_{\alpha }\Phi \nabla^{\alpha }\Phi &\rightarrow& H_{\alpha \beta \gamma } H_{\epsilon \delta 
\varepsilon } H^{\epsilon \delta \varepsilon } \beta^{\beta 
\gamma } \nabla^{\alpha }\Phi - 3( {V_1}^{\beta \gamma } {W_2}_{\beta 
\gamma }- {V_2}^{\beta \gamma } {W_1}_{\beta 
\gamma }  )\nabla_{\alpha }\Phi \nabla^{\alpha }\Phi \,,\nn\\
H_{\alpha }{}^{\gamma \delta
} H_{\beta \gamma \delta } \nabla^{\alpha }\Phi \nabla^{\beta
}\Phi &\rightarrow&  H_{\alpha }{}^{\epsilon \delta } H_{\beta \gamma 
}{}^{\varepsilon } H_{\epsilon \delta \varepsilon } \beta 
^{\beta \gamma } \nabla^{\alpha }\Phi - 2( {V_1}_{\alpha 
}{}^{\gamma } {W_2}_{\beta \gamma }- {V_2}_{\alpha 
}{}^{\gamma } {W_1}_{\beta \gamma }) \nabla^{\alpha }\Phi \nabla^{
\beta }\Phi\,, \nn\\
\nabla_{\alpha }\Phi \nabla^{\alpha }\Phi \nabla_{\beta
}\nabla^{\beta }\Phi &\rightarrow& 2 H_{\beta \gamma \delta } \beta^{\gamma \delta } 
\nabla_{\alpha }\Phi \nabla^{\alpha }\Phi \nabla^{\beta 
}\Phi \,.
\eeqa
Similarly, one can use the transformations \reef{b-trans} and the constraint \reef{b-cons} to determine the transformations of the other couplings in \reef{L1bulk}. In these relations, the 2-forms are defined as $W_1=db_1$, $W_2=db_2$, $V_1=dg_1$, and $V_2=dg_2$, where the vectors are defined as $b_{1\mu}=B_{\mu\nu}\beta_1^\nu$, $b_{2\mu}=B_{\mu\nu}\beta_2^\nu$, $g_{1\mu}=G_{\mu\nu}\beta_1^\nu$, and $g_{2\mu}=G_{\mu\nu}\beta_2^\nu$. Note that the right-hand side terms are covariant and invariant under gauge transformations. By using these transformations, one can determine $\delta(\!\!\bS^1)$.

The $\beta$-constraint requires that the sum of $\delta(\!\!\bS^1)$ and $\alpha'\Delta(\!\!\bS^0)$ must be some total derivative terms. Therefore, we include all possible covariant total derivative terms in our calculations. The most general total derivative terms that are invariant under standard diffeomorphisms and $B$-field gauge transformations involving the spacetime fields and the parameter $\beta$ are given by:
\beqa
\cJ^{(1)}&=&-\frac{2\alpha'}{\kappa^2}\int d^D x\sqrt{-G} \nabla_{\alpha}\Big[e^{-2\Phi}I^{(1)\alpha}\Big],\labell{tot}
\eeqa
where the vector $I^{(1)\alpha}$ is obtained by contracting $R$, $H$, $\nabla\Phi$, $\beta$, and their covariant derivatives at order $\alpha'$. It contains 48 terms
\beqa
I^{(1)\alpha} &\!\!\!\!=\!\!\!\!&j_{1}   H^{\gamma  \delta  \epsilon  } R_{\beta  
\gamma  \delta  \epsilon  } \beta^{\alpha  \beta  } +   
 j_{2}   H^{\alpha  }{}_{\gamma  \delta  } 
R^{\beta  \gamma  } \beta_{\beta  }{}^{\delta  } +  
  j_{9}   H^{\alpha  \delta  \epsilon  } H_{\beta  
\delta  }{}^{\varepsilon  } H_{\gamma  \epsilon  \varepsilon  } 
\beta^{\beta  \gamma  } +   
 j_{10}   H^{\alpha  \delta  \epsilon  } H_{\beta  
\gamma  }{}^{\varepsilon  } H_{\delta  \epsilon  \varepsilon  } 
\beta^{\beta  \gamma  } \nn\\&&+   
 j_{11}   H^{\alpha  }{}_{\beta  }{}^{\delta  } H_{\gamma 
 }{}^{\epsilon  \varepsilon  } H_{\delta  \epsilon  \varepsilon  
} \beta^{\beta  \gamma  } +   
 j_{12}   H^{\alpha  }{}_{\beta  \gamma  } H_{\delta  
\epsilon  \varepsilon  } H^{\delta  \epsilon  \varepsilon  } 
\beta^{\beta  \gamma  } +   
 j_{3}   H^{\alpha  }{}_{\beta  \gamma  } R 
\beta^{\beta  \gamma  } +   
 j_{4}   H_{\beta  }{}^{\delta  \epsilon  } 
R^{\alpha  }{}_{\gamma  \delta  \epsilon  } \beta ^{\beta  \gamma  } \nn\\&&+   
 j_{5}   H_{\beta  }{}^{\delta  \epsilon  } 
R^{\alpha  }{}_{\delta  \gamma  \epsilon  } \beta ^{\beta  \gamma  } +   
 j_{6}   H^{\alpha  \delta  \epsilon  } R_{\beta 
 \gamma  \delta  \epsilon  } \beta^{\beta  \gamma  } +   
 j_{7}   H^{\alpha  \delta  \epsilon  } R_{\beta 
 \delta  \gamma  \epsilon  } \beta^{\beta  \gamma  } +   
 j_{8}   H_{\beta  \gamma  \delta  } R^{\alpha  
\beta  } \beta^{\gamma  \delta  } \nn\\&&+   
 j_{13}   \beta^{\beta  \gamma  } \nabla^{\alpha  
}\nabla_{\delta  }H_{\beta  \gamma  }{}^{\delta  } +   
 j_{19}   H_{\beta  \gamma  \delta  } \nabla^{\alpha  
}\nabla^{\delta  }\beta^{\beta  \gamma  } +   
 j_{20}   H^{\alpha  }{}_{\gamma  \delta  } 
\nabla_{\beta  }\nabla^{\delta  }\beta^{\beta  \gamma  } +  
 j_{
   31}   \beta^{\gamma  \delta  } \nabla^{\alpha  
}H_{\beta  \gamma  \delta  } \nabla^{\beta  }\Phi \nn\\&& +   
 j_{37}   H_{\beta  \gamma  \delta  } \nabla^{\alpha  
}\beta^{\gamma  \delta  } \nabla^{\beta  }\Phi  +   
 j_{43}   H_{\beta  \gamma  \delta  } \beta^{\gamma  
\delta  } \nabla^{\alpha  }\Phi  \nabla^{\beta  }\Phi  +   
 j_{32}   \beta^{\gamma  \delta  } \nabla_{\beta  }H^{
\alpha  }{}_{\gamma  \delta  } \nabla^{\beta  }\Phi  +   
 j_{38}   H^{\alpha  }{}_{\gamma  \delta  } 
\nabla_{\beta  }\beta^{\gamma  \delta  } \nabla^{\beta  
}\Phi  \nn\\&&+   
 j_{44}   H^{\alpha  }{}_{\gamma  \delta  } \beta ^{\gamma  \delta  } \nabla_{\beta  }\Phi  \nabla^{\beta  
}\Phi  +   
 j_{39}   H^{\alpha  }{}_{\beta  \delta  } \nabla^{\beta 
 }\Phi  \nabla_{\gamma  }\beta^{\gamma  \delta  } +   
 j_{14}   \beta^{\beta  \gamma  } \nabla_{\gamma  
}\nabla_{\delta  }H^{\alpha  }{}_{\beta  }{}^{\delta  } +   
 j_{45}   H^{\alpha  }{}_{\gamma  \delta  } \beta _{\beta  }{}^{\delta  } \nabla^{\beta  }\Phi  \nabla^{\gamma  
}\Phi \nn\\&& +   
 j_{33}   \beta^{\gamma  \delta  } \nabla^{\beta  
}\Phi  \nabla_{\delta  }H^{\alpha  }{}_{\beta  \gamma  } +   
 j_{25}   \nabla_{\beta  }\beta^{\beta  \gamma  } 
\nabla_{\delta  }H^{\alpha  }{}_{\gamma  }{}^{\delta  } +   
 j_{34}   \beta_{\beta  }{}^{\gamma  } \nabla^{\beta  
}\Phi  \nabla_{\delta  }H^{\alpha  }{}_{\gamma  }{}^{\delta  } 
+  j_{26}   \nabla^{\alpha  }\beta^{\beta  \gamma  } 
\nabla_{\delta  }H_{\beta  \gamma  }{}^{\delta  } \nn\\&&+   
 j_{35}   \beta^{\beta  \gamma  } \nabla^{\alpha  
}\Phi  \nabla_{\delta  }H_{\beta  \gamma  }{}^{\delta  } +   
 j_{36}   \beta^{\alpha  \gamma  } \nabla^{\beta  
}\Phi  \nabla_{\delta  }H_{\beta  \gamma  }{}^{\delta  } +   
 j_{27}   \nabla^{\gamma  }\beta^{\alpha  \beta  } 
\nabla_{\delta  }H_{\beta  \gamma  }{}^{\delta  } +   
 j_{15}   \beta^{\beta  \gamma  } \nabla_{\delta  
}\nabla^{\alpha  }H_{\beta  \gamma  }{}^{\delta  } \nn\\&&+   
 j_{16}   \beta^{\beta  \gamma  } \nabla_{\delta  
}\nabla_{\gamma  }H^{\alpha  }{}_{\beta  }{}^{\delta  } +   
 j_{17}   \beta^{\alpha  \beta  } \nabla_{\delta  
}\nabla_{\gamma  }H_{\beta  }{}^{\gamma  \delta  } +   
 j_{18}   \beta^{\beta  \gamma  } \nabla_{\delta  
}\nabla^{\delta  }H^{\alpha  }{}_{\beta  \gamma  } +   
 j_{21}   H^{\alpha  }{}_{\beta  \gamma  } 
\nabla_{\delta  }\nabla^{\delta  }\beta^{\beta  \gamma  } \nn\\&&+ 
  j_{46}   H^{\alpha  }{}_{\beta  \gamma  } \beta^{\beta  
\gamma  } \nabla_{\delta  }\nabla^{\delta  }\Phi  +   
 j_{40}   H_{\beta  \gamma  \delta  } \nabla^{\beta  
}\Phi  \nabla^{\delta  }\beta^{\alpha  \gamma  } +   
 j_{41}   H^{\alpha  }{}_{\gamma  \delta  } 
\nabla^{\beta  }\Phi  \nabla^{\delta  }\beta_{\beta  
}{}^{\gamma  } +   
 j_{28}   \nabla^{\alpha  }H_{\beta  \gamma  \delta  } 
\nabla^{\delta  }\beta^{\beta  \gamma  } \nn\\&&+   
 j_{42}   H_{\beta  \gamma  \delta  } \nabla^{\alpha  
}\Phi  \nabla^{\delta  }\beta^{\beta  \gamma  } +   
 j_{29}   \nabla_{\gamma  }H^{\alpha  }{}_{\beta  \delta 
 } \nabla^{\delta  }\beta^{\beta  \gamma  } +   
 j_{30}   \nabla_{\delta  }H^{\alpha  }{}_{\beta  \gamma 
 } \nabla^{\delta  }\beta^{\beta  \gamma  } +   
 j_{22}   H_{\beta  \gamma  \delta  } \nabla^{\delta  
}\nabla^{\alpha  }\beta^{\beta  \gamma  } \nn\\&&+   
 j_{47}   H_{\beta  \gamma  \delta  } \beta^{\beta  
\gamma  } \nabla^{\delta  }\nabla^{\alpha  }\Phi  +   
 j_{23}   H^{\alpha  }{}_{\gamma  \delta  } 
\nabla^{\delta  }\nabla_{\beta  }\beta^{\beta  \gamma  } +  
 j_{48}   H^{\alpha  }{}_{\gamma  \delta  } \beta^{\beta  
\gamma  } \nabla^{\delta  }\nabla_{\beta  }\Phi  +   
 j_{24}   H_{\beta  \gamma  \delta  } \nabla^{\delta  
}\nabla^{\gamma  }\beta^{\alpha  \beta  }\,,\labell{totd}
\eeqa
where $j_1,\cdots, j_{48}$ are some parameters. 

The $\beta$-constraint on the bulk action dictates that the following relation must be satisfied:
\beqa
\delta(\!\!\bS^1)+\alpha'\Delta(\!\!\bS^0)+\cJ^{(1)}&=&0\labell{cons1}.
\eeqa
To solve this equation, we must impose the following Bianchi identities:
\beqa
 R_{\alpha[\beta\gamma\delta]}&=&0\,,\nn\\
 \nabla_{[\mu}R_{\alpha\beta]\gamma\delta}&=&0\,,\labell{bian}\\
\nabla_{[\mu}H_{\alpha\beta\gamma]}&=&0\,,\nn\\
{[}\nabla,\nabla{]}\mathcal{O}-R\mathcal{O}&=&0\,,\nn
\eeqa
as well as the constraint \reef{b-cons}. To impose these constraints, we express the curvatures and covariant derivatives in \reef{cons1} in terms of partial derivatives of the metric and write the field strength $H$ in terms of the $B$-field. This automatically satisfies all the Bianchi identities. Then, we can write the equation \reef{cons1} in terms of independent and non-covariant terms. The coefficients of the independent terms must be zero, which produces some algebraic equations involving all parameters that can be easily solved.

%If the parameters in the total derivative terms and $\beta$-deformations are only required to satisfy the bulk constraint \reef{cons1}, then we can remove any arbitrary unfixed parameters as follows: by excluding $\delta(\!\!\bS^1)$ from \reef{cons1}, the resulting solution provides 38 relations between the parameters in \reef{delta1} and the total derivative terms \reef{totd}, i.e.,
%\beqa
%e_{18}&=&2e_1-e_{15}/2\nn\\
%e_2&=&e_{10}+e_{11}+e_{13}/2+e_{14}/2-e_{17}/2\nn\\ \vdots\nn\\
%j_8&=&-25 e_1/2-e_{13}/2-e_{14}/2-25 e_{15}/8+e_{17}/2+j_{13}\nn\\
%j_9&=&0
%\eeqa
%The relations obtained in the previous step indicate that, as anticipated before, some of the parameters in \reef{Ds0} are related by the Bianchi identities and the total derivative terms, while others are arbitrary and could be fixed by other constraints that may exist for open spacetime manifolds not considered here. In fact, the constraint \reef{cons1} only requires 38 parameters in \reef{delta1} and \reef{totd}. There are many ways to choose these 38 parameters, but the non-zero parameters must be selected such that when they are replaced into \reef{cons1} without $\delta(\!\!\bS^1)$, 38 relations are obtained with right-hand-sides equal to zero. We set all parameters appearing on the right-hand side of these relations to zero, which is one way to keep the 38 parameters in \reef{delta1} and \reef{totd}. Then, we solve \reef{cons1} again by including $\delta(\!\!\bS^1)$.

Interestingly, we obtain the following seven relations between the parameters of the effective action \reef{L1bulk}:
\beqa
a_{3 }&=& a_{1}/3 - a_{10}/8 - (5 a_{12})/48 - (5 a_{14})/384 - (5 a_{15})/144 - (
  25 a_{17})/2304 + a_{18}/72 + a_{19}/576 \nn\\&&+ a_{2}/9 - (5 a_{20})/
  36, \nn\\ a_{4 }&=& -12 a_{1} - a_{15}/4 - a_{16}/16 + a_{19}/16 - 4 a_{2 }- 
  a_{20}, \nn\\
   a_{5 }&=& a_{15} + a_{16}/4 - a_{19}/4 + 4 a_{20},\nn\\
    a_{6 }&= &
 a_{10}/2 - a_{11}/8 + a_{12}/4 - (5 a_{13})/48 + a_{14}/12 - a_{16}/48 + (
  5 a_{17})/96,\nn\\
   a_{7 }&=& a_{11}/2 + a_{13}/4 - a_{14}/8 - a_{17}/16,\nn\\
    a_{8 }&=& 
 24 a_{1},\nn\\
  a_{9 }&=& -12 a_{1}\,.\labell{aaa}
\eeqa
These relations are identical to the ones obtained by imposing the Buscher rules \cite{Garousi:2019wgz}. Note that $a_1$ must be non-zero to be consistent with S-matrix elements. This is because upon substituting the above relations into \reef{L1bulk}, we find that $a_1$ is the coefficient of the Riemann squared term, which must be non-zero for consistency. The parameters in the deformations \reef{deform} and the total derivative terms in \reef{tot} are not entirely determined solely by the parameters $a_1, a_{2}, a_{10}, \cdots, a_{20}$. We have  obtained the following deformations for the $\beta$-transformations:
\beqa
\delta\Phi^{(1)}&\!\!\!\!\!=\!\!\!\!\!&\frac{1}{8} (192   a_{1}  -192   
 a_{2}   + 52   a_{11}   + 52   
 a_{13}   + 12   a_{14}   + 188   
 a_{15}   + 13   a_{16}   \nn\\&& + 24   
 a_{17}  -24   a_{18}  -   
 a_{19}  + 816   
 a_{20}  ) H_{\alpha  \beta  \gamma  } \beta^{\beta  \gamma  } \nabla^{\alpha  }\Phi  + \frac{1}{16} 
(-576   a_{1}   + 64   a_{2}   -156  
  a_{11}  \nn\\&&-152   a_{13}   + 64   
 a_{14}  -204   a_{15}  -39   
 a_{16}  + 26   a_{17}   + 8   
 a_{18}   + 3   a_{19}  -816   
 a_{20}  ) \beta^{\alpha  \beta  } 
\nabla_{\gamma  }H_{\alpha  \beta  }{}^{\gamma  } \nn\\&&+ 
\frac{1}{32} (1344   a_{1}   + 960   
 a_{2}  -624   a_{10} -260   
 a_{11}   -600   a_{12}  -250   
 a_{13}   + 50   a_{14} \nn\\&& -460   
 a_{15}  -65   a_{16}   + 96   
 a_{18}   + 19   a_{19}  -1840   
 a_{20}  ) H_{\alpha  \beta  \gamma  } 
\nabla^{\gamma  }\beta^{\alpha  \beta  }+\cdots,\nn\\
\delta G_{\alpha\beta}^{(1)}&\!\!\!\!\!=\!\!\!\!\!&(-24   a_{1}   -8   
 a_{2}   - \frac{1}{2}   
 a_{15}   - \frac{1}{8}   a_{16}   -   
 a_{18}   + \frac{1}{8}   a_{19}   + 2   
 a_{20}  ) (\beta^{\gamma  \delta  } 
\nabla_{\alpha  }H_{\beta  \gamma  \delta  } + \beta^{\gamma 
 \delta  } \nabla_{\beta  }H_{\alpha  \gamma  \delta  
}) \nn\\&&+ (-12   a_{1}   -4   
 a_{2}   - \frac{3}{4}   
 a_{15}   - \frac{3}{16}   a_{16}   -   
 a_{18}   + \frac{3}{16}   a_{19}   +   
 a_{20}  ) (H_{\beta  \gamma  \delta  } 
\nabla_{\alpha  }\beta^{\gamma  \delta  } + H_{\alpha  
\gamma  \delta  } \nabla_{\beta  }\beta^{\gamma  \delta  }
)  \nn\\&&+ (-2   a_{15}   -8   
 a_{20}  ) (H_{\beta  \gamma  \delta  } 
\beta^{\gamma  \delta  } \nabla_{\alpha  }\Phi  + H_{\alpha  
\gamma  \delta  } \beta^{\gamma  \delta  } \nabla_{\beta  
}\Phi )   \nn\\&&+ (  a_{11}   +   
 a_{13}   + \frac{1}{4}   a_{14}   + 4   
 a_{15}   + \frac{1}{4}   
 a_{16}  + \frac{1}{2}   a_{17}   + 16   
 a_{20}  ) H_{\gamma  \delta  \epsilon  } 
G_{\alpha  \beta  } \beta^{\delta  \epsilon  } 
\nabla^{\gamma  }\Phi    \nn\\&&+ (-96   a_{1}   + 32  
  a_{2}   + 4   a_{18}   -8   
 a_{20}  ) (H_{\beta  \gamma  \delta  } 
\beta_{\alpha  }{}^{\delta  } \nabla^{\gamma  }\Phi  + 
H_{\alpha  \gamma  \delta  } \beta_{\beta  }{}^{\delta  } 
\nabla^{\gamma  }\Phi )   \nn\\&&+ (-72   
 a_{1}   + 8   a_{2}   - \frac{1}{2}   
 a_{15}   - \frac{1}{8}   a_{16}   +   
 a_{18}   + \frac{1}{8}   a_{19}   -2   
 a_{20}  ) (\beta_{\beta  }{}^{\gamma  } 
\nabla_{\delta  }H_{\alpha  \gamma  }{}^{\delta  } + \beta_{\alpha  }{}^{\gamma  } \nabla_{\delta  }H_{\beta  \gamma  
}{}^{\delta  })   \nn\\&&+ (-72   a_{1}   + 8   
 a_{2}   - \frac{1}{2}   
 a_{15}   - \frac{1}{8}   
 a_{16}   + \frac{1}{8}   a_{19}   -2   
 a_{20}  ) (H_{\beta  \gamma  \delta  } 
\nabla^{\delta  }\beta_{\alpha  }{}^{\gamma  } + H_{\alpha  
\gamma  \delta  } \nabla^{\delta  }\beta_{\beta  }{}^{\gamma 
 })   \nn\\&&+ (- \frac{3}{2}   
 a_{11}   - \frac{3}{2}   
 a_{13}   + \frac{5}{8}   a_{14}   -2   
 a_{15}   - \frac{3}{8}   
 a_{16}   + \frac{1}{4}   a_{17}   -8   
 a_{20}  ) G_{\alpha  \beta  } \beta^{\gamma  
\delta  } \nabla_{\epsilon  }H_{\gamma  \delta  }{}^{\epsilon  
}   \nn\\&&+ (12   a_{1}   + 4   a_{2}   -3   
 a_{10}   - \frac{5}{4}   a_{11}   -3   
 a_{12}   - \frac{5}{4}   
 a_{13}   + \frac{1}{4}   
 a_{14}    \nn\\&& - \frac{9}{4}   
 a_{15}   - \frac{5}{16}   
 a_{16}   + \frac{1}{2}   
 a_{18}   + \frac{1}{16}   a_{19}   -9   
 a_{20}  ) H_{\gamma  \delta  \epsilon  } 
G_{\alpha  \beta  } \nabla^{\epsilon  }\beta^{\gamma  \delta 
 }+\cdots,\nn\\
 \delta B_{\alpha\beta}^{(1)}&\!\!\!\!\!=\!\!\!\!\!& (-4   a_{1}   - \frac{4}{3}   
 a_{2}   +   a_{10}   + \frac{1}{4}   
 a_{11}   +   a_{12}   + \frac{1}{4}   
 a_{13}   + \frac{1}{24}   
 a_{14}   + \frac{3}{4}   
 a_{15}  \nn\\&&  + \frac{1}{16}   
 a_{16}   + \frac{1}{12}   
 a_{17}   - \frac{1}{6}   
 a_{18}  - \frac{1}{48}   a_{19}   + 3   
 a_{20}  ) H_{\gamma  \delta  \epsilon  } 
H^{\gamma  \delta  \epsilon  } \beta_{\alpha  \beta  } \nn\\&&+ 
(48   a_{1}   + 16   a_{2}   +   
 a_{15}   + \frac{1}{4}   a_{16}   + 2   
 a_{18}   - \frac{1}{4}   a_{19}   -4   
 a_{20}  ) (R_{\beta  }{}^{\gamma  
} \beta_{\alpha  \gamma  } - R_{\alpha  }{}^{\gamma 
 } \beta_{\beta  \gamma  }) \nn\\&&+ (24   
 a_{1}   -8   a_{2}   - \frac{1}{2}   
 a_{18}   + 2   
 a_{20}  ) (H_{\beta  }{}^{\delta  \epsilon 
 } H_{\gamma  \delta  \epsilon  } \beta_{\alpha  }{}^{\gamma  
} - H_{\alpha  }{}^{\delta  \epsilon  } H_{\gamma  \delta  
\epsilon  } \beta_{\beta  }{}^{\gamma  }) + 48   
 a_{1}   H_{\alpha  \gamma  }{}^{\epsilon  } H_{\beta  
\delta  \epsilon  } \beta^{\gamma  \delta  }\nn\\&& + 192   
 a_{1}   R_{\alpha  \gamma  \beta  \delta  } 
\beta^{\gamma  \delta  } + (-2   
 a_{11}   -2   a_{13}   + \frac{3}{2}   
 a_{14}   - \frac{1}{2}   a_{16}   +   
 a_{17}  ) \beta_{\alpha  \beta  } 
\nabla_{\gamma  }\nabla^{\gamma  }\Phi+\cdots,
\eeqa
where dots represent terms in the deformation in which their coefficients are not in terms of the parameters $a_1, a_2, a_{10}, \dots, a_{20}$. The obtained solution also fixes the total derivative terms \reef{totd}, which we are not concerned with for closed spacetime manifolds.

Upon substituting the relations \reef{aaa} into \reef{L1bulk}, the resulting couplings give the effective action of bosonic string theory if we choose $a_1=1/96$. There are 12 arbitrary parameters $a_2, a_{10},a_{11},\cdots, a_{20}$ that reflect the freedom of the action to use arbitrary covariant and gauge invariant field redefinitions in closed spacetime manifolds. For any choice of these parameters, the effective action appears in a specific scheme. For example, if we set $a_2=-3a_1$ and all other arbitrary parameters to zero, then we obtain the action in the Metsaev-Tseytlin scheme \cite{Metsaev:1987zx}, i.e.,
 \beqa
 \bS^1&=&-\frac{\alpha' }{2\kappa^2}\int d^{26}x\, e^{-2\Phi}\sqrt{-G}\Big[   R_{\alpha \beta \gamma \delta} R^{\alpha \beta \gamma \delta} -\frac{1}{2}H_{\alpha}{}^{\delta \epsilon} H^{\alpha \beta \gamma} R_{\beta  \gamma \delta\epsilon}\nn\\
&&\qquad\qquad\qquad\qquad\qquad\quad+\frac{1}{24}H_{\epsilon\delta \zeta}H^{\epsilon}{}_{\alpha}{}^{\beta}H^{\delta}{}_{\beta}{}^{\gamma}H^{\zeta}{}_{\gamma}{}^{\alpha}-\frac{1}{8}H_{\alpha \beta}{}^{\delta} H^{\alpha \beta \gamma} H_{\gamma}{}^{\epsilon \zeta} H_{\delta \epsilon \zeta}\Big]\,.\labell{S1bf}
\eeqa
 The corresponding $\beta$-transformations are
 \beqa
\delta\Phi^{(1)}&\!\!\!\!\!=\!\!\!\!\!& H_{\alpha  \beta  \gamma  } \beta^{\beta  \gamma  } \nabla^{\alpha  }\Phi - \frac{1}{2} 
 \beta^{\alpha  \beta  } 
\nabla_{\gamma  }H_{\alpha  \beta  }{}^{\gamma  }-
\frac{1}{2} H_{\alpha  \beta  \gamma  } 
\nabla^{\gamma  }\beta^{\alpha  \beta  }+\cdots,\nn\\
\delta G_{\alpha\beta}^{(1)}&=&-2(H_{\beta  \gamma  \delta  } 
\beta_{\alpha  }{}^{\delta  } \nabla^{\gamma  }\Phi  + 
H_{\alpha  \gamma  \delta  } \beta_{\beta  }{}^{\delta  } 
\nabla^{\gamma  }\Phi )  -(\beta_{\beta  }{}^{\gamma  } 
\nabla_{\delta  }H_{\alpha  \gamma  }{}^{\delta  } + \beta_{\alpha  }{}^{\gamma  } \nabla_{\delta  }H_{\beta  \gamma  
}{}^{\delta  })   \nn\\&& -(H_{\beta  \gamma  \delta  } 
\nabla^{\delta  }\beta_{\alpha  }{}^{\gamma  } + H_{\alpha  
\gamma  \delta  } \nabla^{\delta  }\beta_{\beta  }{}^{\gamma 
 }) +\cdots,\nn\\
 \delta B_{\alpha\beta}^{(1)}&=&\frac{1}{2}(H_{\beta  }{}^{\delta  \epsilon 
 } H_{\gamma  \delta  \epsilon  } \beta_{\alpha  }{}^{\gamma  
} - H_{\alpha  }{}^{\delta  \epsilon  } H_{\gamma  \delta  
\epsilon  } \beta_{\beta  }{}^{\gamma  }) + \frac{1}{2}  H_{\alpha  \gamma  }{}^{\epsilon  } H_{\beta  
\delta  \epsilon  } \beta^{\gamma  \delta  }+2  R_{\alpha  \gamma  \beta  \delta  } 
\beta^{\gamma  \delta  }+\cdots ,\labell{MT}
\eeqa
where dots represent terms in the deformation  which have unfixed  parameters. %The action \reef{S1bf} is also invariant under the $\MZ_2$-transformations of T-duality. However, in this case, the corresponding deformed Buscher rules involve the deformation of the base space metric \cite{Garousi:2019wgz}, which spoils the data on the boundary. Consequently, the boundary terms cannot be represented in the Metsaev-Tseytlin scheme such that the combination of bulk and boundary actions satisfies the $\MZ_2$-symmetry.

If we choose the arbitrary parameters as
\beqa
&&a_{10}= -16 a_{1}, a_{11}= 0, a_{12}= 16 a_{1}, a_{13}= 192 a_{1}, a_{14}= 384 a_{1}, a_{15}= 0, \nn\\&&
a_{16}= -384 a_{1}, a_{17}= -384 a_{1}, a_{18}= 48 a_{1}, a_{19}= 0, a_{2}= -3a_{1}, a_{20}= 0,
\eeqa
then we obtain the action in the Meissner scheme \cite{Meissner:1996sa}, i.e.,
\beqa
\bS^1&=&-\frac{\alpha'}{2\kappa^2}\int_M d^{26} x\sqrt{-G} e^{-2\Phi}\Big[R_{GB}^2+\frac{1}{24} H_{\alpha }{}^{\delta \epsilon } H^{\alpha \beta
\gamma } H_{\beta \delta }{}^{\varepsilon } H_{\gamma \epsilon
\varepsilon } -  \frac{1}{8} H_{\alpha \beta }{}^{\delta }
H^{\alpha \beta \gamma } H_{\gamma }{}^{\epsilon \varepsilon }
H_{\delta \epsilon \varepsilon }\nn\\&&  + \frac{1}{144} H_{\alpha
\beta \gamma } H^{\alpha \beta \gamma } H_{\delta \epsilon
\varepsilon } H^{\delta \epsilon \varepsilon }+ H_{\alpha }{}^{
\gamma \delta } H_{\beta \gamma \delta } R^{\alpha
\beta } -  \frac{1}{6} H_{\alpha \beta \gamma } H^{\alpha \beta
\gamma } R  -
\frac{1}{2} H_{\alpha }{}^{\delta \epsilon } H^{\alpha \beta
\gamma } R_{\beta \gamma \delta \epsilon }\nn\\&& -
\frac{2}{3} H_{\beta \gamma \delta } H^{\beta \gamma \delta }
\nabla_{\alpha }\nabla^{\alpha }\Phi + \frac{2}{3} H_{\beta
\gamma \delta } H^{\beta \gamma \delta } \nabla_{\alpha }\Phi
\nabla^{\alpha }\Phi + 8 R \nabla_{\alpha }\Phi
\nabla^{\alpha }\Phi + 16 \nabla_{\alpha }\Phi \nabla^{\alpha
}\Phi \nabla_{\beta }\nabla^{\beta }\Phi \nn\\&&- 16
R_{\alpha \beta } \nabla^{\alpha }\Phi \nabla^{\beta
}\Phi - 16 \nabla_{\alpha }\Phi \nabla^{\alpha }\Phi \nabla_{
\beta }\Phi \nabla^{\beta }\Phi + 2 H_{\alpha }{}^{\gamma
\delta } H_{\beta \gamma \delta } \nabla^{\beta
}\nabla^{\alpha }\Phi \Big]\,,\labell{Mis}
\eeqa
where $R^2_{
GB}$  is the Gauss-Bonnet  couplings. The corresponding $\beta$-transformations are
\beqa
\delta\Phi^{(1)}&\!\!\!\!\!=\!\!\!\!\!&- \frac{1}{8} H_{\alpha \beta \gamma } \nabla^{\gamma }\beta ^{\alpha \beta }+\cdots,\nn\\
\delta G_{\alpha\beta}^{(1)}&=&\frac{1}{4} (H_{\beta \gamma \delta } \nabla_{\alpha }\beta^{\gamma \delta } + H_{\alpha \gamma \delta } \nabla_{\beta }
\beta^{\gamma \delta }) - \frac{1}{2} (H_{\beta \gamma 
\delta } \nabla^{\delta }\beta_{\alpha }{}^{\gamma }+  
H_{\alpha \gamma \delta } \nabla^{\delta }\beta_{\beta }{}^{
\gamma })+\cdots,\nn\\
\delta B_{\alpha\beta}^{(1)}&=&\frac{1}{4} (H_{\beta }{}^{\delta \epsilon } H_{\gamma \delta 
\epsilon } \beta_{\alpha }{}^{\gamma } -  H_{\alpha }{}^{\delta 
\epsilon } H_{\gamma \delta \epsilon } \beta_{\beta }{}^{\gamma }) 
+ \frac{1}{2} H_{\alpha \gamma }{}^{\epsilon } H_{\beta \delta 
\epsilon } \beta^{\gamma \delta } + 2 R_{\alpha \gamma 
\beta \delta } \beta^{\gamma \delta }+\cdots,\labell{ss}
\eeqa
where dots represent terms in the deformation that have unfixed parameters. These parameters may be fixed by requiring the deformed $\beta$-transformations, in combination with the local transformations, to close an algebra. Since all the parameters in the effective actions are already completely fixed, we are not interested in this algebra.
\section{Discussion}

In this paper, we have demonstrated that the standard bulk and boundary NS-NS couplings at the leading order of $\alpha'$ are invariant under the global $\beta$-transformations \reef{b-trans}. We have then investigated this symmetry at order $\alpha'$. Through explicit calculations at this order, we have shown that the effective actions of bosonic string theory in the Metsaev-Tseytlin and Meissner schemes can be obtained by imposing this symmetry.
Furthermore, we have discovered the corresponding deformations of the $\beta$-transformations up to some unfixed parameters, which may be determined by requiring that the combinations of the deformed $\beta$-transformation and local transformations satisfy a closed algebra \cite{Baron:2022but}.

%and observed that they alter the data on the boundary at order $\alpha'$. Therefore, the combination of the bulk and boundary actions at order $\alpha'$ that is explicitly invariant under the $\beta$-transformation may not be representable in covariant schemes. Instead, they may be found in schemes where the local transformations are deformed.

We have observed that without requiring the deformed $\beta$-transformations and local transformations to satisfy a closed algebra, the constraint of invariance of the effective action under the deformed $\beta$-transformations completely fixes all independent couplings at order $\alpha'$, up to field redefinitions. If one starts with all independent couplings, including the most general higher-derivative field redefinitions, the $\beta$-symmetry can fix the couplings up to one overall factor. However, this is not the case for higher orders of $\alpha'$. For example, at order $\alpha'^2$, there are 60 independent couplings up to field redefinitions \cite{Garousi:2019cdn}.
We have conducted calculations at order $\alpha'^2$ and found that the $\beta$-constraint on the effective action results in only 28 relations among the 60 independent parameters at this order. The remaining parameters need to be determined by imposing the condition that the deformed $\beta$-transformations, in combination with the local transformations, satisfy a closed algebra. These 28 relations are consistent with the 59 relations fixed by the Buscher rules. Similarly, in the case of deformed Buscher rules, the invariance of the 60 couplings under these rules generates only some relations among the couplings. The remaining parameters are fixed by requiring that the deformed Buscher rules satisfy the $\mathbb{Z}_2$ group. It would be intriguing to investigate whether imposing the closure algebra and the invariance of the effective action under the $\beta$-transformations can completely determine all 60 parameters of the effective action, up to one overall factor.

{\bf Declaration of generative AI and AI-assisted technologies in the writing process}:
During the preparation of this work, the author utilized Poe to enhance the writing. After utilizing this tool/service, the author reviewed and edited the content as necessary, taking full responsibility for the publication's content.


\begin{thebibliography}{9}

	%\cite{Giveon:1994fu}
\bibitem{Giveon:1994fu}
  A.~Giveon, M.~Porrati and E.~Rabinovici,
  %``Target space duality in string theory,''
  Phys.\ Rept.\  {\bf 244}, 77 (1994)
  doi:10.1016/0370-1573(94)90070-1
  [hep-th/9401139].
  %%CITATION = doi:10.1016/0370-1573(94)90070-1;%%
  %839 citations counted in INSPIRE as of 14 Jan 2017
	%\cite{Alvarez:1994dn}
\bibitem{Alvarez:1994dn}
  E.~Alvarez, L.~Alvarez-Gaume and Y.~Lozano,
  %``An Introduction to T duality in string theory,''
  Nucl.\ Phys.\ Proc.\ Suppl.\  {\bf 41}, 1 (1995)
  doi:10.1016/0920-5632(95)00429-D
  [hep-th/9410237].
  %%CITATION = doi:10.1016/0920-5632(95)00429-D;%%
  %186 citations counted in INSPIRE as of 14 Jan 2017

  %\cite{Sen:1991zi}
\bibitem{Sen:1991zi}
A.~Sen,
%``O(d) x O(d) symmetry of the space of cosmological solutions in string theory, scale factor duality and two-dimensional black holes,''
Phys. Lett. B \textbf{271}, 295-300 (1991)
doi:10.1016/0370-2693(91)90090-D
%206 citations counted in INSPIRE as of 11 May 2020

%\cite{Hohm:2014sxa}
\bibitem{Hohm:2014sxa}
O.~Hohm, A.~Sen and B.~Zwiebach,
%``Heterotic Effective Action and Duality Symmetries Revisited,''
JHEP \textbf{02}, 079 (2015)
doi:10.1007/JHEP02(2015)079
[arXiv:1411.5696 [hep-th]].
%30 citations counted in INSPIRE as of 12 Dec 2020

%\cite{Bergshoeff:1996cy}
\bibitem{Bergshoeff:1996cy}
E.~Bergshoeff and M.~De Roo,
%``D-branes and T duality,''
Phys. Lett. B \textbf{380}, 265-272 (1996)
doi:10.1016/0370-2693(96)00523-0
[arXiv:hep-th/9603123 [hep-th]].
%112 citations counted in INSPIRE as of 02 Feb 2023



%\cite{Garousi:2019wgz}
\bibitem{Garousi:2019wgz}
  M.~R.~Garousi,
  %``Four-derivative couplings via the $T$-duality invariance constraint,''
  Phys.\ Rev.\ D {\bf 99}, no. 12, 126005 (2019)
  doi:10.1103/PhysRevD.99.126005
  [arXiv:1904.11282 [hep-th]].
  %%CITATION = doi:10.1103/PhysRevD.99.126005;%%
  %1 citations counted in INSPIRE as of 14 Jul 2019

 %\cite{Garousi:2019mca}
\bibitem{Garousi:2019mca}
M.~R.~Garousi,
%``Effective action of bosonic string theory at order $\alpha'^2 $,''
Eur. Phys. J. C \textbf{79}, no.10, 827 (2019)
doi:10.1140/epjc/s10052-019-7357-4
[arXiv:1907.06500 [hep-th]].
%12 citations counted in INSPIRE as of 08 Aug 2021

%\cite{Garousi:2019xlf}
\bibitem{Garousi:2019xlf}
M.~R.~Garousi,
%``Surface terms in the effective actions via duality constraints,''
Phys. Lett. B \textbf{809}, 135733 (2020)
doi:10.1016/j.physletb.2020.135733
[arXiv:1907.09168 [hep-th]].
%10 citations counted in INSPIRE as of 07 Oct 2022

   %\cite{Razaghian:2018svg}
\bibitem{Razaghian:2018svg}
  H.~Razaghian and M.~R.~Garousi,
  %``$R^4 $ terms in supergravities via T-duality constraint,''
  Phys.\ Rev.\ D {\bf 97}, 106013 (2018)
  doi:10.1103/PhysRevD.97.106013
  [arXiv:1801.06834 [hep-th]].
  %%CITATION = doi:10.1103/PhysRevD.97.106013;%%


%\cite{Garousi:2020gio}
\bibitem{Garousi:2020gio}
M.~R.~Garousi,
%``Effective action of type II superstring theories at order $\alpha'^3$: NS-NS couplings,''
JHEP \textbf{02}, 157 (2021)
doi:10.1007/JHEP02(2021)157
[arXiv:2011.02753 [hep-th]].
%5 citations counted in INSPIRE as of 02 Mar 2021

%\cite{Garousi:2020lof}
\bibitem{Garousi:2020lof}
M.~R.~Garousi,
%``On NS-NS couplings at order \ensuremath{\alpha}'3,''
Nucl. Phys. B \textbf{971}, 115510 (2021)
doi:10.1016/j.nuclphysb.2021.115510
[arXiv:2012.15091 [hep-th]].
%7 citations counted in INSPIRE as of 22 Nov 2021

%\cite{Garousi:2021yyd}
\bibitem{Garousi:2021yyd}
M.~R.~Garousi,
%``Higher-derivative field redefinitions in the presence of boundary,''
Eur. Phys. J. C \textbf{82}, no.7, 645 (2022)
doi:10.1140/epjc/s10052-022-10611-7
[arXiv:2111.10987 [hep-th]].
%1 citations counted in INSPIRE as of 06 Oct 2022

%\cite{Garousi:2021cfc}
\bibitem{Garousi:2021cfc}
M.~R.~Garousi,
%``Effective action of string theory at order $\alpha '$ in the presence of boundary,''
Eur. Phys. J. C \textbf{81}, no.12, 1149 (2021)
doi:10.1140/epjc/s10052-021-09952-6
[arXiv:2103.13682 [hep-th]].
%4 citations counted in INSPIRE as of 06 Oct 2022

%\cite{Garousi:2022ovo}
\bibitem{Garousi:2022ovo}
M.~R.~Garousi,
%``Background independence of effective actions at critical dimension,''
Phys. Rev. D \textbf{105}, no.10, 106021 (2022)
doi:10.1103/PhysRevD.105.106021
[arXiv:2201.00588 [hep-th]].
%1 citations counted in INSPIRE as of 06 Oct 2022

%\cite{Garousi:2022ghs}
\bibitem{Garousi:2022ghs}
M.~R.~Garousi,
%``Higher-derivative couplings and torsional Riemann curvature,''
JHEP \textbf{12}, 139 (2022)
doi:10.1007/JHEP12(2022)139
[arXiv:2210.17069 [hep-th]].
%1 citations counted in INSPIRE as of 02 Feb 2023




%\cite{Garousi:2013gea}
\bibitem{Garousi:2013gea}
M.~R.~Garousi, A.~Ghodsi, T.~Houri and G.~Jafari,
%``T-duality of D-brane action at order $\alpha'$ in bosonic string theory,''
JHEP \textbf{10}, 103 (2013)
doi:10.1007/JHEP10(2013)103
[arXiv:1308.4609 [hep-th]].
%12 citations counted in INSPIRE as of 06 Oct 2022

  %\cite{Robbins:2014ara}
\bibitem{Robbins:2014ara}
  D.~Robbins and Z.~Wang,
  %``Higher Derivative Corrections to O-plane Actions: NS-NS Sector,''
  JHEP {\bf 1405}, 072 (2014)
  doi:10.1007/JHEP05(2014)072
  [arXiv:1401.4180 [hep-th]].
  %%CITATION = doi:10.1007/JHEP05(2014)072;%%
  %13 citations counted in INSPIRE as of 29 Feb 2020

  %\cite{Garousi:2014oya}
\bibitem{Garousi:2014oya}
  M.~R.~Garousi,
  %``T-duality of O-plane action at order $ \alpha'^2$,''
  Phys.\ Lett.\ B {\bf 747}, 53 (2015)
  doi:10.1016/j.physletb.2015.05.049
  [arXiv:1412.8131 [hep-th]].
  %%CITATION = doi:10.1016/j.physletb.2015.05.049;%%
  %9 citations counted in INSPIRE as of 29 Feb 2020

  %\cite{Akou:2020mxx}
\bibitem{Akou:2020mxx}
Y.~Akou and M.~R.~Garousi,
%``Surface terms in effective action of O-plane at order $\alpha '^2$,''
Eur. Phys. J. C \textbf{81}, no.3, 201 (2021)
doi:10.1140/epjc/s10052-021-08990-4
[arXiv:2012.13264 [hep-th]].
%3 citations counted in INSPIRE as of 17 Jul 2022

%\cite{Mashhadi:2020mzf}
\bibitem{Mashhadi:2020mzf}
M.~Mashhadi and M.~R.~Garousi,
%``O-plane couplings at order $\alpha′^{2}$: one R-R field strength,''
JHEP \textbf{06}, 171 (2020)
doi:10.1007/JHEP06(2020)171
[arXiv:2003.05359 [hep-th]].
%3 citations counted in INSPIRE as of 17 Jul 2022

%\cite{Hosseini:2022vrr}
\bibitem{Hosseini:2022vrr}
M.~R.~Hosseini and M.~R.~Garousi,
%``T-duality of D-brane versus O-plane actions,''
JHEP \textbf{01}, 012 (2023)
doi:10.1007/JHEP01(2023)012
[arXiv:2210.03972 [hep-th]].
%2 citations counted in INSPIRE as of 02 Feb 2023

%\cite{Buscher:1987sk}
\bibitem{Buscher:1987sk}
  T.~H.~Buscher,
  %``A Symmetry of the String Background Field Equations,''
  Phys.\ Lett.\ B {\bf 194}, 59 (1987).
  doi:10.1016/0370-2693(87)90769-6
  %%CITATION = doi:10.1016/0370-2693(87)90769-6;%%
  %689 citations counted in INSPIRE as of 14 Jan 2017
	%\cite{Buscher:1987qj}
\bibitem{Buscher:1987qj}
  T.~H.~Buscher,
  %``Path Integral Derivation of Quantum Duality in Nonlinear Sigma Models,''
  Phys.\ Lett.\ B {\bf 201}, 466 (1988).
  doi:10.1016/0370-2693(88)90602-8
  %%CITATION = doi:10.1016/0370-2693(88)90602-8;%%
  %736 citations counted in INSPIRE as of 14 Jan 2017
  
  
  %\cite{Kaloper:1997ux}
\bibitem{Kaloper:1997ux}
  N.~Kaloper and K.~A.~Meissner,
  %``Duality beyond the first loop,''
  Phys.\ Rev.\ D {\bf 56}, 7940 (1997)
  doi:10.1103/PhysRevD.56.7940
  [hep-th/9705193].
  %%CITATION = doi:10.1103/PhysRevD.56.7940;%%
  %56 citations counted in INSPIRE as of 14 Jan 2017
  
  %\cite{Maharana:1992my}
\bibitem{Maharana:1992my}
J.~Maharana and J.~H.~Schwarz,
%``Noncompact symmetries in string theory,''
Nucl. Phys. B \textbf{390}, 3-32 (1993)
doi:10.1016/0550-3213(93)90387-5
[arXiv:hep-th/9207016 [hep-th]].
%487 citations counted in INSPIRE as of 07 Oct 2020


%\cite{Baron:2022but}
\bibitem{Baron:2022but}
W.~H.~Baron, D.~Marques and C.~A.~Nunez,
%``\ensuremath{\beta} Symmetry of Supergravity,''
Phys. Rev. Lett. \textbf{130}, no.6, 061601 (2023)
doi:10.1103/PhysRevLett.130.061601
[arXiv:2209.02079 [hep-th]].
%5 citations counted in INSPIRE as of 26 Aug 2023

%\cite{Andriot:2014uda}
\bibitem{Andriot:2014uda}
 D.~Andriot and A.~Betz,
%``NS-branes, source corrected Bianchi identities, and more on backgrounds with non-geometric fluxes,''
JHEP \textbf{07}, 059 (2014)
doi:10.1007/JHEP07(2014)059
[arXiv:1402.5972 [hep-th]].
%74 citations counted in INSPIRE as of 26 Apr 2023




%\cite{Marques:2015vua}
\bibitem{Marques:2015vua}
D.~Marques and C.~A.~Nunez,
%``T-duality and \ensuremath{\alpha}'-corrections,''
JHEP \textbf{10}, 084 (2015)
doi:10.1007/JHEP10(2015)084
[arXiv:1507.00652 [hep-th]].
%97 citations counted in INSPIRE as of 10 Dec 2022







%\cite{Gibbons:1976ue}
\bibitem{Gibbons:1976ue}
  G.~W.~Gibbons and S.~W.~Hawking,
  %``Action Integrals and Partition Functions in Quantum Gravity,''
  Phys.\ Rev.\ D {\bf 15}, 2752 (1977).
  doi:10.1103/PhysRevD.15.2752
  %%CITATION = doi:10.1103/PhysRevD.15.2752;%%
  %2360 citations counted in INSPIRE as of 21 Jul 2019
  
  %\cite{Metsaev:1987zx}
\bibitem{Metsaev:1987zx}
  R.~R.~Metsaev and A.~A.~Tseytlin,
  %``Order alpha-prime (Two Loop) Equivalence of the String Equations of Motion and the Sigma Model Weyl Invariance Conditions: Dependence on the Dilaton and the Antisymmetric Tensor,''
  Nucl.\ Phys.\ B {\bf 293}, 385 (1987).
  doi:10.1016/0550-3213(87)90077-0
  %%CITATION = doi:10.1016/0550-3213(87)90077-0;%%
  %378 citations counted in INSPIRE as of 04 Sep 2017
	
 %\cite{Meissner:1996sa}
\bibitem{Meissner:1996sa}
K.~A.~Meissner,
%``Symmetries of higher order string gravity actions,''
Phys. Lett. B \textbf{392}, 298-304 (1997)
doi:10.1016/S0370-2693(96)01556-0
[arXiv:hep-th/9610131 [hep-th]].
%120 citations counted in INSPIRE as of 07 Oct 2020

%\cite{Nutma:2013zea}
\bibitem{Nutma:2013zea}
  T.~Nutma,
  %``xTras : A field-theory inspired xAct  package for mathematica,''
  Comput.\ Phys.\ Commun.\  {\bf 185}, 1719 (2014)
  doi:10.1016/j.cpc.2014.02.006
  [arXiv:1308.3493 [cs.SC]].
  %%CITATION = doi:10.1016/j.cpc.2014.02.006;%%
  %49 citations counted in INSPIRE as of 16 Jan 2018
  
  %\cite{Garousi:2019cdn}
\bibitem{Garousi:2019cdn}
M.~R.~Garousi and H.~Razaghian,
%``Minimal independent couplings at order $\alpha'^2$,''
Phys. Rev. D \textbf{100}, no.10, 106007 (2019)
doi:10.1103/PhysRevD.100.106007
[arXiv:1905.10800 [hep-th]].
%13 citations counted in INSPIRE as of 01 Jul 2023

	
\end{thebibliography}
\end{document}